\documentclass[apjl]{emulateapj}
\def\um{\hbox{\,$\mu$m}}
\def\hr{\hbox{$^\mathrm{h}$}}
\def\m{\hbox{$^\mathrm{m}$}}
\def\s{\hbox{$^\mathrm{s}$}}
\def\pm{\hbox{$\pm$}}
\def\n2h{\hbox{N$_2$H$^+$}}

\def\h13co{\hbox{H$^{13}$CO$^{+}$}}

\slugcomment{}

\shorttitle{Primordial substructure in NGC\,2264}
\shortauthors{Teixeira et al.}

\begin{document}


\title{Identifying Primordial Substructure in NGC\,2264}

\author{Paula S. Teixeira\altaffilmark{1,2}, Charles J. Lada\altaffilmark{1}, Erick T. Young\altaffilmark{3}, Massimo Marengo\altaffilmark{1}, August Muench\altaffilmark{1}, James Muzerolle\altaffilmark{3}, Nick Siegler\altaffilmark{3}, George Rieke\altaffilmark{3}, Lee Hartmann\altaffilmark{4}, S. Thomas Megeath\altaffilmark{1}, and Giovanni Fazio\altaffilmark{1}}
\altaffiltext{1}{Harvard-Smithsonian Center for Astrophysics, 60 Garden Street, Cambridge, MA 02138, USA; pteixeira@cfa.harvard.edu, clada@cfa.harvard.edu, mmarengo@cfa.harvard.edu, gmuench@cfa.harvard.edu, tmegeath@cfa.harvard.edu, gfazio@cfa.harvard.edu.}
\altaffiltext{2}{Departamento de F\'{\i}sica da Faculdade de Ci\^encias da \hbox{Universidade} de Lisboa, Portugal}
\altaffiltext{3}{Steward Observatory, University of Arizona, 933 North Cherry Avenue, Tucson, AZ 85721, USA; \hbox{eyoung@as.arizona.edu}, jamesm@as.arizona.edu, nsiegler@as.arizona.edu, grieke@as.arizona.edu.}
\altaffiltext{4}{Dept. Astronomy, University of Michigan, 500 Church St.,830 Dennison Building, Ann Arbor, MI   48109, lhartm@umich.edu.}

\begin{abstract}
We present new \emph{Spitzer} Space Telescope observations of the young cluster NGC\,2264. Observations at 24\um\ with the Multiband Imaging Photometer has enabled us to identify the most highly embedded and youngest objects in NGC\,2264. This letter reports on one particular region of NGC\,2264 where  bright 24\um\ sources are spatially configured in curious linear structures with quasi-uniform separations.
 The majority of these sources ($\sim$\,60\%) are found to be  protostellar in nature with Class I spectral energy distributions. Comparison of their spatial distribution with sub-millimeter data from \citet{wolf-chase} and millimeter data from \citet{peretto05} shows a close correlation between the dust filaments and the linear spatial  configurations of the protostars, indicating that star formation is occurring primarily within dense dusty filaments. Finally, the quasi-uniform separations of the protostars are found to be comparable in magnitude to the expected Jeans length suggesting thermal fragmentation of the dense filamentary material.
\end{abstract}


%---------------------------------------------------------------------
\section{Introduction}
\label{sec:introduction}
%---------------------------------------------------------------------

NGC\,2264 is an extended hierarchically structured young cluster \citep{ladas03} associated with a giant molecular cloud in the Monoceros OB1 complex and located at a distance of 800\,pc \citep[e.g.][]{dahm05}. The cluster is a very well stu\-died region \citep[e.g.][]{herbig54,walker54}, displaying evidence of ongoing star formation such as a plethora of molecular outflows \citep{margulis88,wolf-chase} and Herbig-Haro objects \citep{reipurth04}. Luminous far infrared sources were detected by the InfraRed Astronomical Satellite (IRAS), many of them being Class 0 and Class I sources \citep{margulis89}. However detailed study of the protostars in NGC\,2264 was hampered by the limiting sensitivity and resolution of IRAS. The \emph{Spitzer} Space Telescope provides significantly higher resolution and sensitivity. We have thus re-visited NGC\,2264 by surveying the cluster with both the Infrared Array Ca\-me\-ra (IRAC) and the Multiband Imaging Photometer for \emph{Spitzer} (MIPS).

 Here we present initial results of a MIPS and IRAC survey of a very young star forming region within NGC\,2264, originally observed in the infrared by \citet{sargent84} and \citet{schwartz85}. This par\-ti\-cu\-lar region has been also observed at other wavelengths, namely in the sub-millimeter \citep{williams02,wolf-chase} and millimeter \citep{peretto05}. These observations reveal several filamentary structures extending from a central IRAS source, \objectname{IRAS\,06382+0939} (also designated as IRAS\,12 by \citet{margulis89} who classified it as a Class I source). IRAS\,12's position appears to be a few arcseconds away from a near-infrared binary discovered by \citet{castelaz88}, RNO-West (B-type star) and RNO-East (low-mass star). \citet{simon05} have photometric and spectroscopic observations of this binary in X-ray and near-infrared wavelengths and propose that RNO-E could be a proto-Herbig Ae star.
Our \emph{Spitzer} data indicate that IRAS\,12 is in fact coincident with this binary.
 Several sub-millimeter cores \citep{williams02,wolf-chase} and millimeter cores \citep{peretto05} are identified within these filaments. We compare results obtained with our \emph{Spitzer} data with these sub-milli\-meter and millimeter observations in \S \ref{sec:results}. Finally, we discuss primordial cluster sub-structuring in this region of NGC\,2264 and compare these results with those found for the Trapezium \citep{lada04} in \S \ref{sec:discussion}.

%---------------------------------------------------------------------
\section{Observations and Data Reduction}
\label{sec:observations}
%---------------------------------------------------------------------

NGC\,2264 was observed with IRAC as part of the \emph{Spitzer} Guaranteed  
Time Observation program 37 \citep{fazio04}. The data were acquired in two  
epochs seven months apart (March 6, 2004 and October 8, 2004), with  
two dithers at each epoch to allow easy removal of asteroids and  
other transients. The total mosaicked area corres\-ponds to $\sim 0.7^ 
\circ \times 1.2^\circ$. The observations were performed in all IRAC  
bands (centered at 3.6, 4.5, 5.8 and 8.0\um) using the 12\,s  
IRAC High Dynamic Range mode, consisting of two consecutive  
exposures with 0.4 and 10.4\,s integration time at each dither  
position. Basic data reduction and calibration were done with the  
\emph{Spitzer} Science Center pipeline, version S10.5. A final mosaic  
was created for each of the two HDR exposures, using the SSC mosaic  
software the MOPEX (version 10/15/04), resampling the individual images  
on a final pixel scale of 0.86267\arcsec/pix ($1/\sqrt{2}$ of the original  
IRAC pixel scale) to have optimal point source registering at all IRAC  
bands. Cosmic rays and other outliers were removed using MOPEX  
temporal outlier module, and the diffuse background emission was  
matched between individual frames using the MOPEX overlap correction  
module.

We generated a list of point sources in each mosaic, at each  
wavelength, using the SExtractor package \citep{sextractor},  
and then performed aperture photometry using the IRAF routine APPHOT  
with an aperture radius of 2 pixels ($\sim 1.73$\arcsec). IRAC band mer\-ging
was performed using the IDL function srcor\footnote{from IDL Astronomy User's Library: \url{http://idlastro.gsfc.nasa.gov/homepage.html}, adapted from software from the Ultraviolet Imaging Telescope.}, where we used a matching radius of 2 pixels. 
The saturated sources had their magnitudes replaced with those obtained by the short
exposure observations (0.4\,s).

The MIPS \citep{Rieke04} observations were conducted on 2004 March
16 using the scan map mode.  Fourteen scan legs of 0.75\arcdeg  \/
length and 160\arcsec \/ offsets were taken at medium speed.  Total
integration times of 80\, s per point and 40\,s per point were
obtained in the 24 $\mu$m band and 70 $\mu$m band, respectively.  We
also obtained sparse coverage in the 160 $\mu$m band, but most of
those data were saturated due to the extremely high backgrounds from the
molecular cloud. The full map was centered at 6\hr40\m55\s\ +9\degr37\arcmin08\arcsec\
with a position angle of 179\arcdeg. These observations were
processed with the MIPS Data Analysis Tool (DAT) \citep{Gordon05}
which produces calibrated mosaics of the mapped regions. Processing
of the resultant image products to obtain photometry was done using
DAOPHOT and IDL routines. We used a zero point of 7.3\,Jy for the 24\um\ magnitude scale.

%---------------------------------------------------------------------
\section{Results and Analysis}
\label{sec:results}
%---------------------------------------------------------------------

\begin{figure}[!h]
\centering
\includegraphics[scale=0.45]{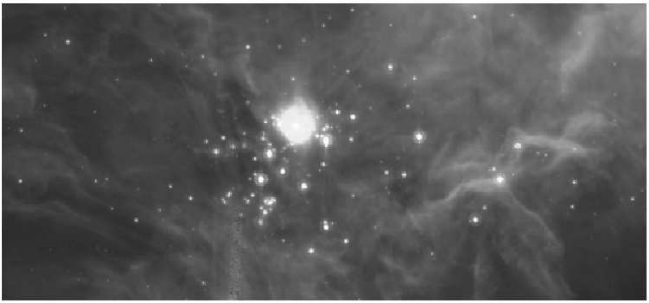}
\caption{False color image of the Spokes cluster constructed from MIPS 24\um\ (red), IRAC 8.0\um\ (green), and IRAC 3.6\um\ (blue) data. The image shows unusual linear spatial alignments of the brightest 24\um\ sources. The central saturated source is IRAS\,06382+0930 (IRAS\,12).}
\label{fig:spokes}
\end{figure}

Figure \ref{fig:spokes} shows a color composite image of the region in NGC\,2264 surrounding IRAS\,12, which is the bright saturated source. The more extended nebulosity (green) in the image corresponds to polycyclic aromatic hydrocarbon (PAH) emission from the molecular cloud at 7.6-8\um\ and detected in the IRAC 8\um\ band. 
The brightest 24\um\ sources (red) form several linear structures that are not clearly evident in the shorter wavelength data. The li\-ne\-ar arrangements of the brightest 24\um\ sources appear as arms extending from the brightest source, IRAS\,12, resembling ``spokes'' of a wheel. We therefore refer to this region as the ``Spokes'' cluster. 
 We performed two simple statistical tests to ascertain whether the linear spatial alignments of sources could occur by chance. The first test consisted of examining 10,000 synthetic fields, with randomly generated distributions of objects. Each field has the same number of objects as the bright 24\um\ sources in the Spokes cluster. These fields were searched for linear alignments that consist of 5 or more stars within a cone of an opening angle $<$ 10\degr\ and a length $<$ 3\arcmin. The second test consisted of radially binning the bright 24\um\ sources, using IRAS\,12 as the center, and comparing the average bin density with a Poissonian distribution. Both tests yield a probability of 0.01-0.02\% of finding an alignment of stars like that observed in the Spokes cluster.

To analyze the nature of the sources in the Spokes cluster we constructed Spectral Energy Distributions (SEDs) for all the sources in the selected field using \emph{Spitzer} data. We determined the spectral indices, $\alpha_\mathrm{IRAC}$, of those sources detected in all four IRAC bands (3.6\um, 4.5\um, 5.8\um, and 8\um) by performing a linear fit to the IRAC flux points in a log$(\lambda)$ vs. log$(\lambda$F$_\lambda)$ diagram, for each source. Identifying Class I objects as sources with \hbox{$\alpha_\mathrm{IRAC} >$ 0} \citep{lada87}, we estimate the total fraction of Class I 24\um\ sources in the entire field to be 33\%.
\begin{figure}[!h]
\centering
\includegraphics[scale=0.45,angle=90]{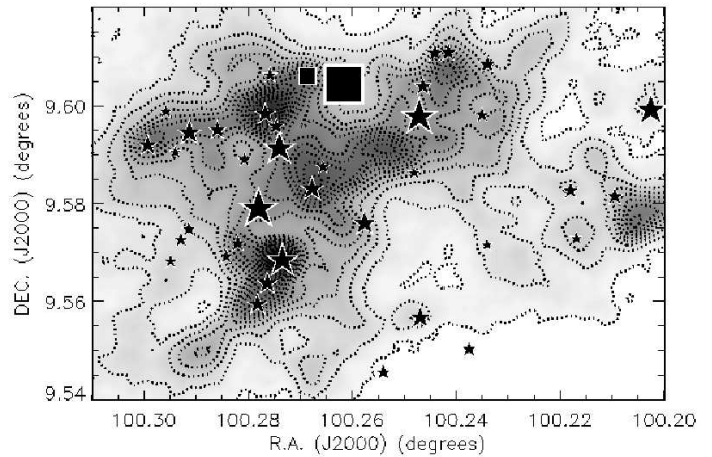}
\caption{Comparison of the spatial locations of 24\um\ MIPS sources with dust emission at 850\um\ (SCUBA data courtesy from \citet{wolf-chase}). The greyscale and contours represent the sub-millimeter dust emission (contours range from 0 to 2 in steps of 0.1 Jy/beam), while the star symbols mark the position of the sources detected at 24\um\ with MIPS. The two squares mark the positions of the two brightest 24\um\ sources ($<$ 2 magnitudes). The sizes of the star and box symbols are proportional to the magnitude of the sources (ranging from 2.0 to 6.0 magnitudes for the star symbols). The beam size for the 850\um\ data is indicated in the lower right corner.}
\label{fig:submm}
\end{figure}
To better comprehend the spatial alignment of these protostars we exa\-mine sub-millimeter observations by \citet{wolf-chase} which trace dust emission from dense molecular gas.
Figure \ref{fig:submm} shows 850\um\ emission detected in the region \citep{wolf-chase}. The star symbols mark the positions of the bright 24\um\ sources and the symbol sizes are proportional to the magnitudes of the sources. It is remarkable how well the bright 24\um\ sources are aligned with the dusty filaments. The same result is obtained when using the 450\um\ data, also from \citet{wolf-chase}. The comparison with the sub-millimeter data shows that the linear alignments have drawn our eye to filaments of the interstellar medium that appear to be breaking up and forming stars in a very regular pattern.

\begin{figure}
\centering
\includegraphics[scale=0.35]{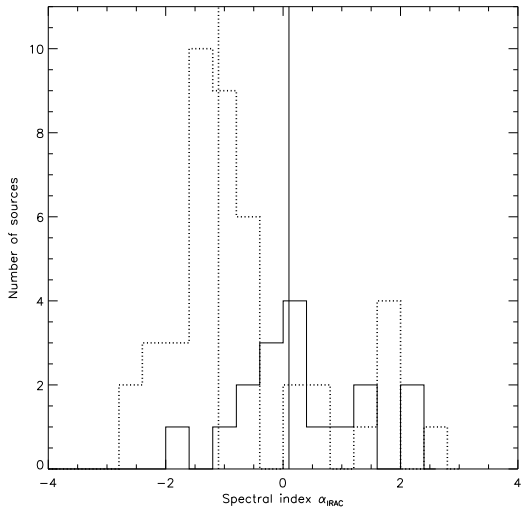}
\caption{Diagram of the distribution of spectral indices for sources detected at 24\um\ within (solid line) the 850\um\ 0.4 Jy/beam contour (see Figure \ref{fig:submm}), and for the remaining sources in the field (dotted line). The vertical solid line corresponds to the median of the distribution of spectral indices of sources inside the contour, $\alpha_\mathrm{IRAC}$=0.1, whereas the vertical dotted line corresponds to the median of the distribution of spectral indices of sources outside the contour, $\alpha_\mathrm{IRAC}$=-1.1.}
\label{fig:histogram}
\end{figure}

We compare the distribution of $\alpha_\mathrm{IRAC}$ for sources within the dusty filaments and for sources in the surroun\-ding region, using the 850\um\ 0.4\,Jy/beam contour in Figure \ref{fig:submm} as the boundary separating these two regions. Figure \ref{fig:histogram} shows a histogram of the spectral indices determined for both the 24\um\ sources within (solid line) and outside (dotted line) this contour. The diagram in Figure \ref{fig:histogram} shows that there is a higher concentration of Class I sources within the dusty filaments, as seen by comparing the median values of $\alpha_\mathrm{IRAC}$ for both distributions: 0.1 for sources inside the filaments and \hbox{-1.1} for the remaining surrounding sources. The fraction of Class I 24\um\ sources within the dusty filaments is 59\% whereas for outside the filaments it is 23\%.  
Not all the 24\um\ sources in the Spokes cluster have detections in all the IRAC bands.
Of these remaining sources that have at least 3 band (IRAC and MIPS) detections we find 7 with rising slopes. 
We list all the Class I sources that we identified within the dusty filaments in Table 1, sorted by their 24\um\ magnitude. The spectral index $\alpha_\mathrm{IRAC}$ is tabulated in column 4. We note that there are two 24\um\ sources within the dusty filaments that have no IRAC counterpart, and as such we were not able to classify them. These sources could also be very young protostars.

Recent millimeter observations by \citet{peretto05} identify 1.2\,mm emission peaks in this region that we find are spatially coincident with 9-10 of the sources listed in Table 1\footnote{specifically D-MM1, D-MM2, D-MM3, D-MM6, D-MM8, D-MM9, D-MM10, D-MM13, and D-MM15  from \citet{peretto05}, corresponding to sources with \# 4, 9, 7, 16, 5, 10, 15, 11, and 2, respectively.}
 (other millimeter sources detected outside the filaments also have IRAC and/or MIPS detections). 
\begin{figure}[!h]
\centering
\includegraphics[scale=0.35]{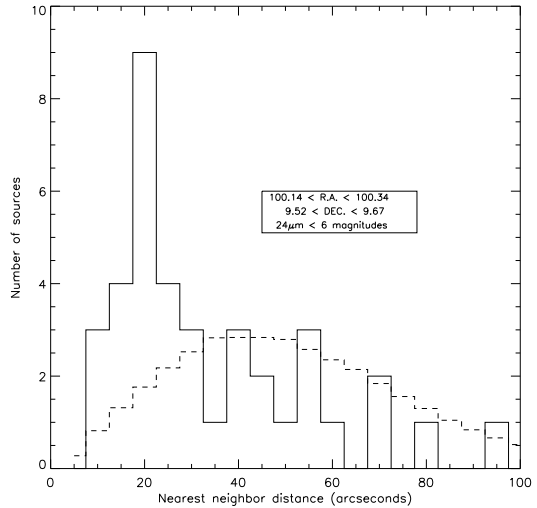}
\caption{Histogram of the nearest neighbour distances between the 24\um\ sources in the Spokes cluster (solid line). We have restricted this sample to sources brighter than the 6th magnitude at 24\um. The dashed line histogram corresponds to the averaged nearest neighbour distance distributions of 10,000 simulated fields populated with randomly positioned stars (see text for details). The preferential separation between the observed 24\um\ sources is \hbox{20\arcsec\ $^+_-$ 5\arcsec}.}
\label{fig:jeanslength}
\end{figure}

Finally, we analyse the spatial separations of the protostars in the Spokes cluster.
Figure \ref{fig:jeanslength} shows a histogram (solid line) of the nearest (projected) neighbor separations between the brightest 24\um\ sources in the Spokes cluster. There is a clear peak in the distribution corresponding to a separation of 20\arcsec. 
The mean separation of sources in this distribution is 26\arcsec. 
By using Monte Carlo techniques, we generated 10,000 synthetic star fields of the same area and number of stars as in the observed field. The dashed-line histogram in Figure \ref{fig:jeanslength} corresponds to the averaged nearest neighbor distributions of these simulated fields. We calculated the probability of generating a field with the same fraction of sources, i.e. 44.7\%, with nearest neighbor separations between 15\arcsec\ and 25\arcsec\ as in the Spokes cluster field. The probability of finding a field with a 40\% or greater fraction of sources with those separations is 0.08\%. The 20\arcsec\ peak is therefore very unlikely a random feature.

\citet{williams02} determined a mean density from their sub-millimeter observations of 3$\times$10$^4$\,cm$^{-3}$ for the dense gas in the Spokes cluster. Assuming a temperature of 17\,K \citep{ward-thompson00} the corresponding Jeans length for this density is 0.105\,pc or 27\arcsec\ at a distance of 800\,pc. For a mean density of 4.2$\times$10$^4$\,cm$^{-3}$ the corresponding Jeans length would be 20\arcsec. These Jeans lengths agree very well with the separations we measure, possibly indicating that the dense filaments have thermally fragmentated into quasi-equidistant cores where the stars are now forming. This result is reminiscent of that found in the Taurus clouds by \citet{hartmann02}.

\begin{deluxetable}{ccccc}[!h]
\scriptsize
\tablecolumns{5}
\tablewidth{0pc}
\tablecaption{Class I sources in the Spokes cluster dusty filaments}
\tablehead{
\colhead{ID}    &  \multicolumn{2}{c}{position (J2000)} & \colhead{$\alpha_\mathrm{IRAC}$} &  \colhead{m$_{[24\um]}$} \\
\colhead{} & \colhead{R.A.} & \colhead{DEC.} & \colhead{}  & \colhead{(mag.)} 
}
\startdata
1   & 06\hr41\m02.8\s & +09\degr36\arcmin16\arcsec  & \nodata & 0.44 \\ % 51
2   & 06\hr41\m04.5\s & +09\degr36\arcmin21\arcsec & 2.33\tablenotemark{a}    & 1.00 \\ % 78
3   & 06\hr41\m05.8\s & +09\degr35\arcmin29\arcsec & 0.52    & 2.20 \\ % 22
4   & 06\hr41\m05.6\s & +09\degr34\arcmin08\arcsec & 0.03\tablenotemark{b}   & 2.21 \\ % 73
5   & 06\hr41\m01.8\s & +09\degr34\arcmin34\arcsec & 0.13   & 2.93  \\ % 16
6   & 06\hr41\m09.9\s & +09\degr35\arcmin41\arcsec & 2.36    & 2.94 \\ % 67
7   & 06\hr41\m04.3\s & +09\degr34\arcmin59\arcsec & 0.09 & 3.07  \\% 63
8   & 06\hr41\m06.3\s & +09\degr33\arcmin50\arcsec & 1.36   &  3.33 \\ % 11
9   & 06\hr41\m06.5\s & +09\degr35\arcmin54\arcsec & 3.26\tablenotemark{b}    & 3.42  \\% 75
10  & 06\hr41\m06.8\s & +09\degr33\arcmin35\arcsec & 1.42\tablenotemark{c}   & 3.54 \\ % 79
11  & 06\hr41\m08.6\s & +09\degr35\arcmin42\arcsec & 1.99\tablenotemark{b}    & 3.87 \\  % 74
12  & 06\hr40\m58.0\s & +09\degr36\arcmin40\arcsec & 1.15   & 4.11 \\ % 69
13  & 06\hr40\m59.1\s & +09\degr36\arcmin15\arcsec & 0.65\tablenotemark{b}    & 4.11 \\ % 76
14  & 06\hr40\m58.6\s & +09\degr36\arcmin39\arcsec & 1.23    & 4.19 \\ % 30
15  & 06\hr41\m07.7\s & +09\degr34\arcmin19\arcsec & 2.22    & 4.29 \\  % 59
16  & 06\hr40\m58.0\s & +09\degr36\arcmin15\arcsec & 0.12   & 6.77 \\ % 68
17  & 06\hr41\m04.5\s & +09\degr33\arcmin44\arcsec & 0.02  & 7.22 \\ % 58
\enddata
\tablenotetext{a}{$\alpha_\mathrm{IRAC}$ determined by fitting 3.6\um, 4.5\um, and 8.0\um}
\tablenotetext{b}{$\alpha_\mathrm{IRAC}$ determined by fitting 4.5\um, 5.8\um, and 8.0\um}
\tablenotetext{c}{$\alpha_\mathrm{IRAC}$ determined by fitting 5.8\um, 8.0\um, and 24\um}
\end{deluxetable}

%---------------------------------------------------------------------
\section{Discussion and Conclusions}
\label{sec:discussion}
%---------------------------------------------------------------------

Our analysis of the \emph{Spitzer} observations of NGC\,2264 leads us to conclude that we are identifying the primordial substructure of this cluster. Bright 24\um\ sources, spatially arranged in unusual linear patterns, were found to be lying along and within dense fingers of molecular material with a spacing similar to that expected for simple Jeans fragmentation. These sources are likely very young if we consider the short crossing times of the associated sub-millimeter and millimeter cores suggested by \citet{williams02} of 0.5\,Myr and \citet{peretto05} of 0.63\,Myr, respectively. In fact, we find the majority ($\sim$60\%) of these sources to be Class I sources and protostellar candidates, suggesting that star formation in the Spokes cluster is occurring primarily within dense filamentary molecular structures.

Primordial filamentary substructuring of this kind has also been found in 
the Trapezium cluster, in Orion. \citet{lada04}, by analyzing deep ground-based 3.4\um\ observations of the Trapezium, find a deeply embedded population of young objects that traces a filamentary molecular ridge lying behind the main cluster. 
 The older, more evolved foreground population in the Trapezium is distributed in a more dispersed manner with an isothermal-like distribution. Similar to the Trapezium, NGC\,2264 also exhibits a somewhat older and more dispersed population near the Spokes cluster \citep{ladas03}. 
\citet{williams02} observed \h13co\ emission lines and \citet{peretto05} observed \n2h\ lines toward the sub-millimeter and millimeter peaks in the Spokes cluster, which they used to measure the core velocity dispersion and infer the virial mass of the system. They find that the virial mass is less than the total gas and dust mass enveloping the cluster. This means that the cores within the dusty filaments are currently bound to the filaments. If the total stellar mass of the cluster is less than the total mass of the dense gas in the filaments then the stellar group is destined to become unbound if the surrounding gas and dust is rapidly dispersed. 
This situation suggests that after star formation occurs in filamentary molecular fragments of the cloud, the more recently formed stars eventually expand to merge with the older population as the cloud material is rapidly cleared through outflows and jets, leading to the less structured distributions of the overall po\-pu\-la\-tion \citep{bonnell03}.

Finally, we find that the bright 24\um\ sources appear to have a preferential separation between themselves of \hbox{20\arcsec\ $^+_-$ 5\arcsec}. This distance is in very good agreement with the Jeans length corresponding to the mean density of the IRS-2 region determined by \citet{williams02}. It appears we are observing the result of the thermal fragmentation of the dense filamentary material into quasi-equidistant star-forming cores.
On the other hand, turbulent motions appear to be present in this region: \citet{peretto05} find that the millimeter peaks have ve\-lo\-ci\-ty dispersions greater than the thermal sound speed for mo\-le\-cu\-lar gas at 15\,K. Perhaps the dusty filaments may have formed through turbulent motions of the cloud \citep{bonnell03}, then as the turbulence decayed the filaments became thermalized and fragmented into star-forming cores. The newly formed stars would generate outflows that re-energize the surrounding region, feeding turbulence back into the cores and filaments and eventually disrupting the cloud. The positions of the young protostars may be the only remaining indication that thermal pressure played an important role in the formation of these objects.

\acknowledgments

We are indebted to Grace Wolf-Chase for kindly sharing with us her sub-millimeter data. 
PT acknowledges support from the scholarship SFRH/BD/13984/2003 awarded by the Funda\c{c}\~ao para a Ci\^encia e Tecnologia (Portugal).
This work is based on observations made with the Spitzer Space Telescope, which is operated by the Jet Propulsion Laboratory, California Institute of Technology under NASA contracts 1407 and 960785.

\end{document}